\documentclass[aps,prb,twocolumn,superscriptaddress,showpacs,longbibliography]{revtex4-1}
\usepackage{amsmath,amssymb,graphicx,subfigure}

\usepackage[utf8]{inputenc}
\usepackage[T1]{fontenc}
\usepackage{xcolor}

\IfFileExists{newtxtext.sty}
   {\usepackage{newtxtext,newtxmath}}
   {\IfFileExists{stix.sty}
      {\usepackage{stix}}
      {\IfFileExists{mathptmx.sty}
      {\usepackage{mathptmx}}{} } }

\usepackage{textcomp}

\usepackage{bm}

\IfFileExists{siunitx.sty}{\usepackage{booktabs,siunitx}}{}

\pdfoutput=1
\usepackage{color}
\definecolor{LinkColor}{rgb}{0.256,0.439,0.588}
\usepackage{hyperref}
\hypersetup{
   colorlinks=true,
   citecolor=LinkColor,
   linkcolor=LinkColor,
   urlcolor=LinkColor
}

\renewcommand{\vec}[1]{\mathbf{#1}}
\newcommand{\bra}[1]{\langle#1\rvert}
\newcommand{\ket}[1]{\lvert#1\rangle}

\newcommand{\be}{\begin{equation}}
\newcommand{\ee}{\end{equation}}
\newcommand{\bea}{\begin{eqnarray}}
\newcommand{\eea}{\end{eqnarray}}

\usepackage{pifont}

\begin{document}

\title{Topological insulators and higher-order topological insulators from gauge-invariant 1D lines}

\author{Heqiu Li}
\affiliation{Department of Physics, University of Michigan, Ann Arbor, MI 48109, USA}
\author{Kai Sun}
\affiliation{Department of Physics, University of Michigan, Ann Arbor, MI 48109, USA}

\begin{abstract}

In this manuscript, we study the interplay between symmetry and topology with a focus on the $Z_2$ topological index of 2D/3D topological insulators and high-order topological insulators.
We show that in the presence of either a two-fold-rotational symmetry or a mirror symmetry, a gauge-invariant quantity can be defined for arbitrary 1D lines in the Brillouin zone.
Such 1D quantities provide a new pathway to compute the $Z_2$ index of topological insulators.
In contrast to the generic setup, where the $Z_2$ index generally involves 2D planes in the Brillouin zone with a globally-defined smooth gauge, this 1D approach only
involves some 1D lines in the Brillouin zone without requiring a global gauge. Such a simplified approach can be used in any time-reversal invariant insulators with
a two-fold crystalline symmetry, which can be found in 30 of the 32 point groups. In addition, this 1D quantity can be further generalized to higher-order topological insulators to compute the magnetoelectric polarization $P_3$. 
\end{abstract}
\date{\today}

\maketitle

\section{Introduction}

In the study of topological states of matter, strong interplays between symmetry and topology have long been known to play an important role. In general, such interplays can be largely classified into two categories: (1) some symmetries are essential for the definition of a topological index, while (2) some other symmetries, although not essential, can provide an easy access to the topological index. For example, the time-reversal symmetry $T$ is essential for the definition of the $Z_2$ index of topological insulators (TIs)~\cite{Kane2005,FuKane_Z2,TI3D,Qi_TI,Hasan_TI,FuKane_inv,Qi2008}. The space-inversion symmetry $I$, in contrast, is not necessary for the $Z_2$ index, but its presence can dramatically simplify the calculation of this index. Without the space-inversion symmetry, it generally requires information about the 2D planes in the Brillouin zone to determine the value of the $Z_2$ index, but in the presence of space inversion, it only requires parity eigenvalues at a few discrete momentum points~\cite{FuKane_inv}. Similarly, the topological monopole charge of topological nodal line semimetals~\cite{Fang2016rev,Bzdusek2017,Ahn2018,Fang:2015aa,ndline,Song2018,Lau2019b} requires the space-time inversion symmetry $IT$, and it in general needs the wave-functions of the Brillouin zone to compute~\cite{Fang:2015aa}. However, when additional (non-essential) rotation symmetries are introduced, the index can be easily computed using rotation eigenvalues at high symmetry points~\cite{ndline,Song2018}, similar to the fact that the Chern number can be computed by rotation eigenvalues~\cite{Fang2012}.
In both examples, non-essential symmetries simplified the topological indices from relying on the entire Brillouin zone to only a few limited momentum points. In particular, because these indices often require a smooth global gauge, which is often challenging to obtain, such symmetry-induced simplification greatly reduces the complexity of the index calculation.

Similar ideas of using symmetries to simplify the calculation of topological indices have been explored extensively in various types of topological states, which have led to many intriguing results~\cite{Fang2012,topchem,indicator,Kruthoff2017,S4index,Ono2018,Song2018b,Zhang2019,Tang2019,Vergniory2019,Alex2016,Lau2016,Miert2017,Kooi2019,Kruthoff2019}. On the practical side, these results provide a highly efficient way to determine the topological properties of various materials and systems, independent of microscopic details of the band structure. In addition, on the fundamental level, these efforts provide a bridge to link various topological indices/phenomena with symmetry representations at high symmetry momenta, such as the indicator theory and topological quantum chemistry~\cite{topchem,indicator,Kruthoff2017,S4index,Ono2018,Song2018b,Zhang2019,Tang2019,Vergniory2019}.

In this study, we focus on systems where symmetries are not strong enough to allow
high symmetry points to fully dictate the topological index, e.g., a $Z_2$ topological insulator with a $C_2$ point group symmetry, and try to understand the role of
non-essential symmetries in these systems.
We show that for an insulator with time-reversal symmetry in 2D or 3D, as long as the system has either a two-fold-rotational symmetry or a mirror symmetry, the $Z_2$ index can be simplified to involve only one-dimensional lines in the Brillouin zone. This result is an extension of the high-symmetry-point-based approaches (e.g. the parity criterion~\cite{FuKane_inv}) from 0D points to 1D lines. Here, the topological index can still be simplified by the non-essential symmetry, although high-symmetry points are no longer sufficient to dictate the index. This simplification is achieved by a gauge-invariant line quantity $g(\widetilde{k_a k_b})$ that we defined below. In the presence of a two-fold-rotational or mirror symmetry, this quantity remains gauge-invariant for any 1D path in the momentum space. By exploring its connection to the Wilson loop approach~\cite{Yu2011,Alex2014}, we show that this line quantity has a physical meaning that it is a measure of the relative phase of Pfaffian in the parallel transport gauge~\cite{smoothgauge}. Furthermore, this quantity also provides a gauge-independent way to calculate the magnetoelectric polarization $P_3$ for higher-order topological insulators~\cite{Benalcazar2017,Benalcazar2017b,Bernevig_HOTI,Schindler2018,pfHOTI,Ahn2019,Geier2018,Tri2019,Ben2019}.

Our result also brings convenience to the calculation of $Z_2$ index in practice, because it only requires some 1D lines in the Brillouin zone to be evaluated, while the original definition of the $Z_2$ index requires a smooth gauge in higher dimensions. And the evaluation can be further simplified into the parity criterion if the space inversion symmetry is present.
This approach applies generically to any time-reversal invariant insulators with a two-fold crystalline symmetry, such as a two-fold-rotation $C_2$, a mirror symmetry or space inversion, which can be found in 30 of all the 32 point groups, with the only exceptions being the $C_3$ group and the trivial group $C_1$. It provides the same level of simplification as the approaches based on the partial polarization~\cite{Kooi2019,Kruthoff2019}, and it can be further generalized to higher-order topological insulators.

\section{Gauge-invariant quantity for arbitrary lines in the Brillouin zone}

\begin{figure}
\includegraphics[width=1.4 in]{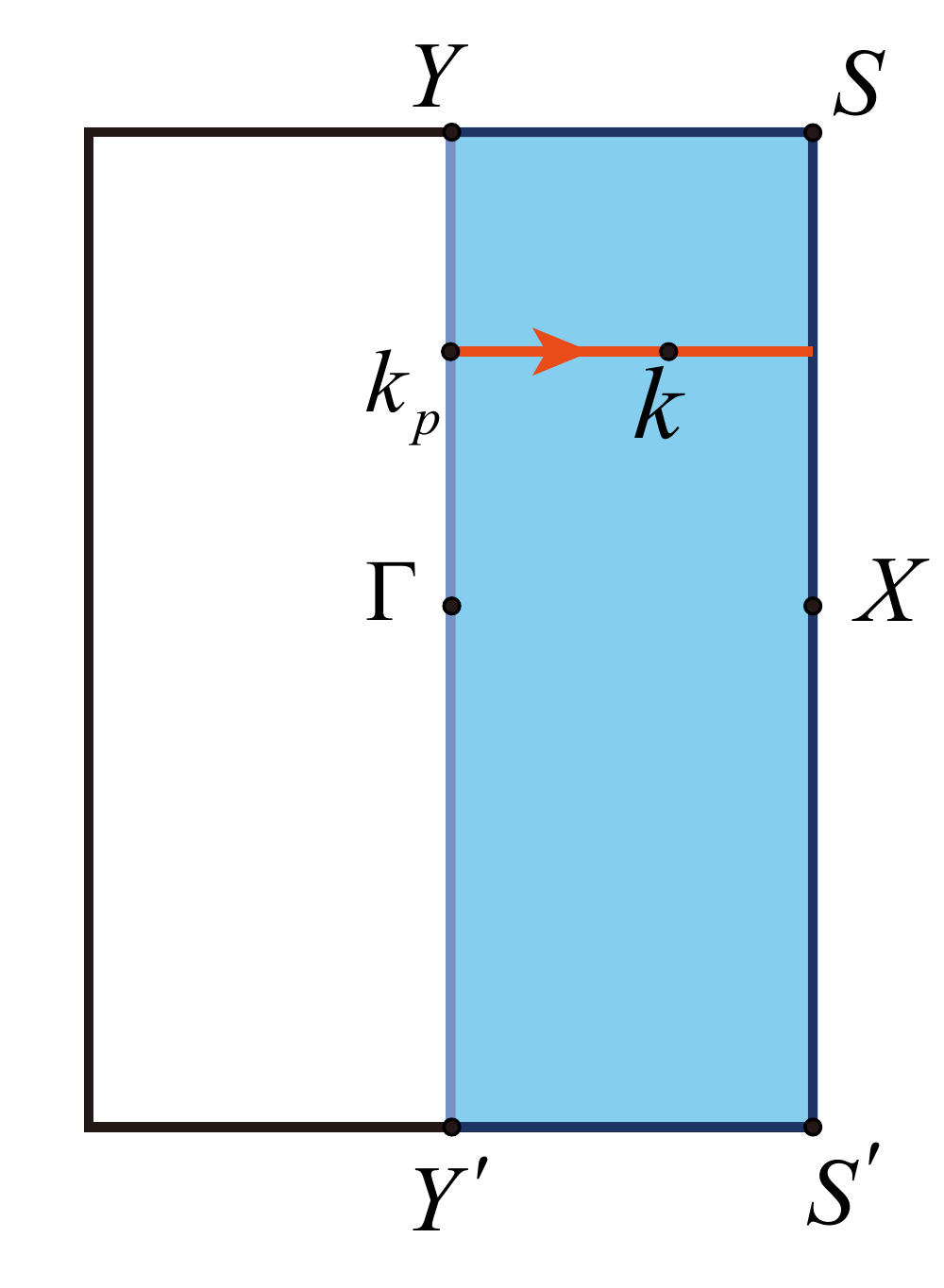}
\caption{ The Brillouin zone for 2D insulator with time reversal symmetry. $\tau$ is represented by the shaded area enclosed by $SYY'S'$.  }
\label{bz2d}
\end{figure}

In this section we define a gauge-invariant line quantity $g(\widetilde{k_a k_b})$ for systems with time-reversal symmetry $T$ and discuss some general properties of it, and in the next section we will show that with additional two-fold crystalline symmetry like two-fold rotation or mirror reflection, this quantity can be utilized to simplify the calculation of the Fu-Kane-Mele (FKM) $Z_2$ index~\cite{Kane2005,FuKane_Z2,TI3D}, although the eigenvalues of the two-fold symmetry are insufficient to determine the index. For an arbitrary path $\widetilde{k_a k_b}$ in the Brillouin zone (BZ) that connects momenta $k_a$ and $k_b$, we can define $g(\widetilde{k_a k_b})$ as:
\bea
g(\widetilde{k_a k_b})&=&\frac{\operatorname{Pf}[M(k_b)]}{\operatorname{Pf}[M(k_a)]}\det[ W(k_a,k_b) ] \label{gwilson}\\
W_{mn}(k_a,k_b)&=&\langle u_m(k_a)| \prod_{k_i\in \widetilde{k_a k_b}}^{k_a\leftarrow k_b} P_{k_i} | u_n(k_b)\rangle \nonumber\\
M_{mn}(k)&=&\bra{u_m(k)} T \ket{u_n(k)} \nonumber
\eea
Here $\operatorname{Pf}$ refers to Pfaffian, $m,n$ refer to occupied bands, $P_k=\sum_{m\in occ} \ket{u_m(k)}\bra{u_m(k)}$ is the projection operator to occupied bands and the product is path-ordered along $\widetilde{k_a k_b}$. The gauge-invariance of $g(k_a,k_b)$ comes from the fact that for a general gauge transformation $|u_m(k)\rangle\rightarrow U_{nm}(k)|u_n(k)\rangle$, $\det[W(k_a,k_b)]$ will be multiplied by $\det[U(k_a)]^*\det[U(k_b)]$, which cancels the change in the Pfaffian. Aside from the Pfaffian part, Eq.\eqref{gwilson} is the determinant of the Wilson line at path $\widetilde{k_a k_b}$. The determinant of a Wilson line is gauge-dependent in general, contrary to a closed Wilson loop~\cite{Yu2011,Alex2014}. The Pfaffian $\operatorname{Pf}[M(k)]$ is gauge-dependent as well, but Eq.\eqref{gwilson} shows the combination of the two is gauge-invariant and thus smooth in any gauge. In the special case when $\widetilde{k_a k_b}$ is taken to be the straight line connecting time-reversal invariant momenta (TRIM), for example $k_a=\Gamma=(0,0,0)$ and $k_b=X=(\pi,0,0)$, Eq.(\ref{gwilson}) coincide with the exponential of the partial polarization $\nu_{1D}$~\cite{FuKane_Z2,Lau2016,Miert2017,Kooi2019,Kruthoff2019}:
\bea
\nu_{1D}&=&\frac{1}{\pi}\left[ \int_\Gamma^X dk \operatorname{Tr} A(k)+i \log \frac{\operatorname{Pf}[\omega(X)]}{\operatorname{Pf}[\omega(\Gamma)]} \right] \mod\ 2 \nonumber\\
&=& \frac{1}{\pi}i\log[g(\overline{\Gamma X})] \mod\ 2
\label{nu1d}
\eea
Here $A(k)$ is the Berry connection and $\omega_{mn}(k)=\bra{u_m(-k)} T \ket{u_n(k)}$. For a general path $\widetilde{k_a k_b}$, $g(\widetilde{k_a k_b})$ has a physical meaning that it is a measure of the relative phase between the Pfaffian at $k_a$ and $k_b$ in the parallel transport gauge. To illustrate this point, we define the parallel transport gauge~\cite{smoothgauge} at path $\widetilde{k_a k_b}$ so that for each $k\in\widetilde{k_a k_b}$,
\be
\ket{u_m(k)}=\prod_{k_i\in \widetilde{k k_a }}^{k\leftarrow k_a} P_{k_i} \ket{u_m(k_a)}
\label{pgauge}
\ee
Here the product is path-ordered from $k_a$ to $k$ along $\widetilde{k_a k_b}$. With this gauge choice along the path, the determinant part in $g(\widetilde{k_a k_b})$ becomes unity, leading to
\be
g(\widetilde{k_a k_b})=\frac{\operatorname{Pf}[M(k_b)]}{\operatorname{Pf}[M(k_a)]}
\label{gratio}
\ee
Hence $g(\widetilde{k_a k_b})$ represents the ratio between the Pfaffian $\operatorname{Pf}[M]$ at $k_b$ and $k_a$ in the parallel transport gauge. This interpretation is useful, because the topological index is related to the winding of the phase of Pfaffian. For example, in a 2D system as shown in Fig.(\ref{bz2d}), the FKM $Z_2$ index is given by~\cite{Kane2005,FuKane_Z2}:
\be
\nu_{2D}=\frac{1}{2\pi i} \oint_{\partial \tau}  dk \cdot \mathbf{\nabla} \log \operatorname{Pf}[M(k)]  \label{pf2D}
\ee
Here $\tau$ is the area enclosed by $SYY'S'$ and $\partial \tau$ is its boundary. The Pfaffian itself is gauge-dependent, making it challenging to track its phase, therefore the evaluation of Eq.(\ref{pf2D}) implicitly requires a smooth gauge on $\tau$. The quantity $g(\widetilde{k_a k_b})$ is naturally gauge-invariant, therefore with the interpretation in Eq.(\ref{gratio}) it provides a convenient way to bypass the gauge issue and track the phase of Pfaffian.

Now we will show that $\nu_{2D}$ can be rewritten in terms of the gauge-invariant quantity $g$. From now on it is sufficient to restrict the path $\widetilde{k_a k_b}$ to be the straight line $\overline{k_a k_b}$ connecting $k_a$ and $k_b$ and define:
\bea
g(k_a, k_b)&=&g(\overline{k_a k_b}) \nonumber\\
&=&\frac{\operatorname{Pf}[M(k_b)]}{\operatorname{Pf}[M(k_a)]} \det \left[ \langle u_m(k_a)| \prod_{k_i\in \overline{k_a k_b}}^{k_a\leftarrow k_b} P_{k_i} | u_n(k_b)\rangle \right]
\label{gwilsonl}
\eea
We construct a parallel transport gauge on $\tau$ as follows~\cite{smoothgauge}. First, find a smooth gauge along straight line $\overline{YY'}$. This can always be done by, for example, parallel transport from $Y'$ upwards to $Y$, then perform a unitary twist along the path $\overline{YY'}$ to recover the periodic condition between $Y$ and $Y'$. Then for each $k\in \tau$, denote its projection to $\overline{YY'}$ as $k_p$ such that $\overline{k_p k}$ is parallel to $\overline{YS}$, as shown in Fig.(\ref{bz2d}). The parallel transport gauge is defined by setting the wave function at $k$ to be:
\be
\ket{u_m(k)}=\prod_{k_i\in \overline{k_p k}}^{k\leftarrow k_p} P_{k_i} \ket{u_m(k_p)}
\ee
The gauge constructed in this way is smooth on $\tau$ and satisfies the periodic boundary condition between $\overline{YS}$ and $\overline{Y'S'}$, because the wave function at $\overline{YS}$ and $\overline{Y'S'}$ are parallel transported from the same set of wave functions at $Y$ and $Y'$. For each $k_p\in\overline{YY'}$, we have $k_p+\pi\hat x\in \overline{SS'}$ and Eq.\eqref{gratio} implies
\be
g(k_p,k_p+\pi\hat x)=\frac{\operatorname{Pf}[M(k_p+\pi\hat x)]}{\operatorname{Pf}[M(k_p)]}
\ee
This relation provides a way to reformulate the $Z_2$ index $\nu_{2D}$ in terms of $g$. Define $\overline{g}(k)=g(k,k+\pi\hat x)$, then Eq.\eqref{pf2D} becomes
\bea
\nu_{2D}&=& \frac{1}{2\pi i} \left(\int_{S'}^S-\int_{Y'}^Y \right)  dk \cdot \mathbf{\nabla} \log \operatorname{Pf}[M(k)] \nonumber\\
&=& \frac{1}{2\pi i} \int_{Y'}^Y   dk \cdot \mathbf{\nabla} \log \frac{\operatorname{Pf}[M(k+\pi\hat x)]}{\operatorname{Pf}[M(k)]} \nonumber\\
&=&\frac{1}{2\pi i} \int_{Y'}^Y   dk \cdot \mathbf{\nabla} \log \overline{g}(k)
\label{gpfaffian}
\eea
Here the final integral is along a time-reversal invariant path that connects $Y,\Gamma$ and $Y'$. Eq.\eqref{gpfaffian} shows that $\nu_{2D}$ can be expressed in terms of the line quantity $g$. Although Eq.\eqref{gpfaffian} is derived in the specific parallel transport gauge, the gauge-invariance of $g$ implies that this formula is valid in any gauge. This means that the above construction of the smooth parallel transport gauge is only a conceptual step which is never needed in a real calculation. Importantly, the evaluation of $g$ itself does not require a smooth gauge to begin with. In practice, to evaluate $\nu_{2D}$ from Eq.\eqref{gpfaffian}, all we need is to make a discrete mesh of $k$ points in $\tau$ with randomly selected phase for each wave function, and by definition in Eq.\eqref{gwilsonl} the function $\overline{g}(k)=g(k,k+\pi\hat x)$ will be smooth in $k$, which can lead to a well-defined result in Eq.\eqref{gpfaffian}. This formalism is important for our later application to higher-order topological insulators. Up to now Eq.\eqref{gpfaffian} requires only time-reversal symmetry, which is the least requirement in symmetry to protect the FKM $Z_2$ topological index.

\section{Topological insulators with two-fold crystalline symmetries in addition to time-reversal symmetry}

We will show that if a 2D or 3D time-reversal invariant topological insulators (TIs) have an additional two-fold crystalline symmetry, for example two-fold rotation $C_2$ or mirror $\sigma$, the line quantity $g$ defined above in Eq.\eqref{gwilson} or Eq.\eqref{gwilsonl} can simplify the FKM $Z_2$ index $\nu$ to involve only one-dimensional lines of the BZ. Note that the eigenvalues of the spacial symmetry $C_2$ or $\sigma$ at high symmetry momenta are not sufficient to determine the $Z_2$ index, because for a spin one-half system $C_2$ or $\sigma$ has eigenvalues $\pm i$ and the two bands in each Kramers pair have opposite eigenvalues of $C_2$ or $\sigma$. Therefore, at any $C_2$ invariant momentum there are always a half of valence bands with $C_2$ or mirror eigenvalue $+i$ and the other half with eigenvalue $-i$ no matter $\nu$ is trivial or not. Hence spacial symmetry eigenvalues themselves are insufficient to determine the topological index.

\subsection{Systems with additional two-fold rotational symmetry}

\begin{figure}
\includegraphics[width=3.4 in]{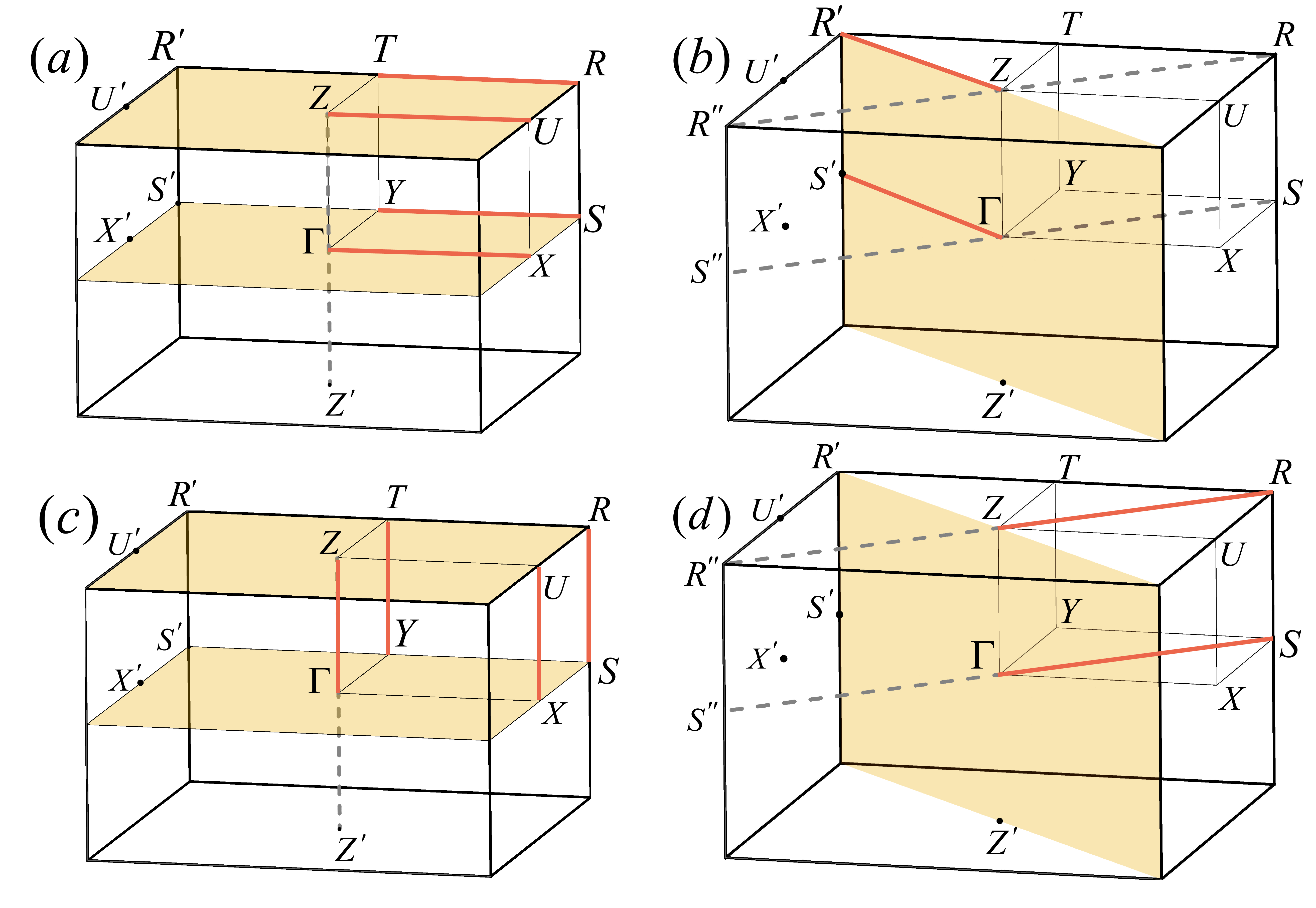}
\caption{ The plot of a 3D Brillouin zone. The red lines denote the one-dimensional subspace that is needed to evaluate the FKM strong $Z_2$ index in Eq.\eqref{gnuc2} and \eqref{gnumir}. (a) System with a $C_2$ symmetry along $\hat z$ direction. The colored planes are the $C_2T$ invariant planes. (b) System with a $C_2$ symmetry along $\hat x+\hat y$ direction. There is only one $C_2T$ invariant plane. (c) System with a mirror plane perpendicular to $\hat z$. (d) System with a mirror plane perpendicular to $\hat x+\hat y$.  }
\label{bzc2}
\end{figure}

When the system has additional spacial symmetries, the line quantity $g(k_a,k_b)$ has a useful property that if a point $k$ in the BZ is transformed to $Ck$ under a symmetry operator $C$ which commutes with time-reversal $T$, then depending on whether $C$ is unitary or anti-unitary, $g(k_a,k_b)$ in Eq.\eqref{gwilsonl} satisfies:
\bea
g(Ck_a,Ck_b)&=&g(k_a,k_b),\ \textmd{if C is unitary } \nonumber \\
g(Ck_a,Ck_b)&=&g(k_a,k_b)^*,\ \textmd{if C is anti-unitary}
\label{gsym}
\eea
This property comes from the transformation properties of the Pfaffian and the Wilson line operator. We give a detailed proof of this property in the appendix. This property is important in evaluating $g(k_a,k_b)$ at symmetry-related lines in the BZ. Now we consider systems with a two-fold rotational symmetry $C_2$ in additional to time-reversal $T$. The combined operator $C_2T$ is also a symmetry of the system which acts on the momentum space like a mirror, therefore there are $C_2T$ invariant planes in the BZ. Suppose there are two such planes that include all eight time-reversal invariant momenta (TRIM) as in Fig.(\ref{bzc2})(a), where the $C_2$ axis is along $z$ direction and $C_2T$ invariant planes are colored in yellow. If $k_a,k_b$ are taken inside one of the $C_2T$ invariant planes, then Eq.\eqref{gsym} shows $g(k_a,k_b)=g(C_2T k_a,C_2T k_b)^*=g(k_a,k_b)^*$ so that $g(k_a,k_b)$ is real because $C_2T$ is anti-unitary. If $k_a,k_b$ are taken to be TRIM, then $g(k_a,k_b)$ should have magnitude of 1 because the $M$ matrices in Eq.\eqref{gwilsonl} are unitary at TRIM. These two conditions quantize $g(K_1,K_2)$ to $\pm 1$, where $K_1,K_2$ are TRIM in the same $C_2T$ invariant plane. Furthermore, as in Eq.\eqref{nu1d}, $g(K_1,K_2)$ is related to the partial polarization $\nu_{1D}$:
\bea
\nu_{1D}(K_1,K_2)&=&\frac{1}{\pi}\left[ \int_{K_1}^{K_2} dk \operatorname{Tr} A(k)+i \log \frac{\operatorname{Pf}[\omega(K_2)]}{\operatorname{Pf}[\omega(K_1)]} \right] \mod\ 2 \nonumber\\
g(K_1,K_2)&=&e^{i\pi \nu_{1D}(K_1,K_2)}=\pm 1
\label{gnu1d}
\eea
It has been shown in Ref.\onlinecite{Kruthoff2019} that the FKM $Z_2$ index for a 2D system is equivalent to the difference between the $1D$ partial polarization $\nu_{1D}$ at two pairs of TRIM:
\bea
\nu_{2D}(k_z=0)&=&\nu_{1D}(\Gamma,X)-\nu_{1D}(Y,S)\ \mod 2 \nonumber\\
\nu_{2D}(k_z=\pi)&=&\nu_{1D}(Z,U)-\nu_{1D}(T,R)\ \mod 2
\label{nu2d}
\eea
The FKM strong $Z_2$ index in 3D is given by $\nu_{3D}=\nu_{2D}(k_z=0)-\nu_{2D}(k_z=\pi)\mod 2$. Combining with Eq.\eqref{gnu1d} we have
\be
(-1)^{\nu_{3D}}=g(\Gamma,X)g(Y,S)g(Z,U)g(T,R)
\label{gnu3da}
\ee
Therefore the 3D strong $Z_2$ index is simplified by the line quantity $g(k_a,k_b)$ so that it involves only 1D subspace of the BZ.

Eq.\eqref{gnu3da} assumes that all the eight TRIM are included in some $C_2T$ invariant planes, which is not always true for a general $C_2$ rotation. For example, if the $C_2$ axis is along direction $\hat x+\hat y$ rather than $\hat z$ for the same Brillouin zone as in Fig.(\ref{bzc2})(b), there is only one $C_2T$ invariant plane $S_1$ passing through $\Gamma,S',R',Z$. Denote the other time-reversal invariant plane passing through $X,Y,T,U$ as $S_2$, Then $\nu_{3D}=\nu_{2D}(S_1)-\nu_{2D}(S_2)\mod 2$. We show that $\nu_{2D}(S_2)$ must vanish due to the $C_2$ symmetry. Denote the midpoint of $\overline{XY}$ as $M$. Since $C_2$ is unitary, from Eq.\eqref{gsym} we have $g(X,M)=g(C_2 X,C_2 M)=g(Y,M)=g(M,Y)^{-1}$. Therefore $g(X,Y)=g(X,M)g(M,Y)=1$, which means the partial polarization $\nu_{1D}(X,Y)$ is quantized to 0. The same argument can be applied to $\overline{TU}$. Hence $\nu_{1D}(X,Y)=\nu_{1D}(T,U)=0$ due to the $C_2$ symmetry, therefore $\nu_{2D}(S_2)$ vanishes. We present a more detailed proof of the triviality of $\nu_{2D}(S_2)$ in the appendix using the interpretation of $g(k_a,k_b)$ as a measure of Pfaffian in the parallel transport gauge. With this result, the strong $Z_2$ index $\nu_{3D}$ is determined only by lines in $C_2T$ invariant plane $S_1$:
\be
(-1)^{\nu_{3D}}=(-1)^{\nu_{2D}(S_1)}=g(\Gamma,S')g(Z,R')
\label{gnu3db}
\ee

The above proof for Eq.\eqref{gnu3da} and \eqref{gnu3db} for different types of $C_2$ axis can be unified in a general framework, which is also applicable to Brillouin zones that are not cube-shaped. There are eight TRIM $K_i,i=1...8$ in a 3D BZ. Since time-reversal operator commutes with $C_2$, $C_2$ must bring one TRIM to itself or to another TRIM. For those TRIM non-invariant under $C_2$, let $C_2 K_i=K_j$ and $C_2 K_j=K_i$, then the midpoint $M=(K_i+K_j)/2$ must be invariant under $C_2$, and $g(K_i,M)=g(C_2 K_i,C_2 M)=g(K_j,M)=g(M,K_j)^{-1}$. Therefore $g(K_i,K_j)=g(K_i,M)g(M,K_j)=1$. This fixes the partial polarization $\nu_{1D}(K_i,K_j)=0$, which does not contribute to the strong index $\nu_{3D}$. Therefore we only need to consider those TRIM that are also invariant under $C_2$. Define the $C_2T$ invariant subspace in the BZ by $S_{C_2T}=\{k\in BZ|C_2Tk=k+G \}$ where $G$ is any reciprocal lattice vector, and define a set $L_{C_2T}$ to be the set of straight paths inside $S_{C_2T}$ such that each path $\gamma\in L_{C_2T}$ connects two $C_2$-invariant TRIM, different paths in $L_{C_2T}$ do not cross with each other, and each $C_2$-invariant TRIM is connected by one path in $L_{C_2T}$. With this definition, in Fig.(\ref{bzc2}) (a) and (b) $L_{C_2T}$ reduces to the red lines $\{\overline{\Gamma X},\overline{YS},\overline{ZU},\overline{TR} \}$ and $\{\overline{\Gamma S'},\overline{ZR'}\}$ respectively (or equivalently $\{\overline{\Gamma Y},\overline{XS},\overline{ZT},\overline{UR} \}$ and $\{\overline{\Gamma Z},\overline{S'R'}\}$, which does not change the final result). Therefore Eq.\eqref{gnu3da} and \eqref{gnu3db} can be unified as
\be
(-1)^{\nu_{3D}}=\prod_{\gamma\in L_{C_2T}}  \textmd{Sign}[g(\gamma)]
\label{gnuc2}
\ee
Here the sign function $\textmd{Sign}$ is added to take account of the fact that the absolute value of $g(\gamma)$ can be smaller than 1 in a practical numerical calculation in which the projection in Eq.\eqref{gwilsonl} is taken at discrete points. In this case $g(\gamma)$ is still real, and the sign of it determines the index $\nu_{3D}$. Eq.\eqref{gnuc2} shows that for any topological insulator with an additional two-fold rotational symmetry, the FKM strong $Z_2$ index can be computed through a well-defined one-dimensional subspace of the Brillouin zone. If there are multiple $C_2$ axes in the system, this simplification can be applied to any one of the $C_2$, independent of the presence of the other symmetries. This calculation is applicable to 2D insulators as well, which can be achieved by restricting $L_{C_2T}$ to the 2D Brillouin zone. This method is convenient to implement in the sense that it does not require a smooth gauge, because the definition of $g(\gamma)$ in Eq.\eqref{gwilsonl} is gauge-independent.

\subsection{Systems with additional mirror symmetry}

For systems with a mirror symmetry $\sigma$ in additional to $T$, similar analysis can be applied by evaluating $g(k_a,k_b)$ in the $\sigma T$ invariant subspace defined by $S_{\sigma T}=\{k\in BZ|\sigma Tk=k+G \}$. If the mirror is perpendicular to $\hat z$ as in Fig.(\ref{bzc2})(c), then $S_{\sigma T}$ consists of four straight paths along $\hat z$ direction that pass through TRIM. Since $\sigma T$ is anti-unitary, Eq.\eqref{gsym} implies $g(\Gamma,Z)=g(\sigma T\Gamma,\sigma T Z)^*=g(\Gamma,Z)^*$, therefore $g(\Gamma,Z)$ is real and quantizes to $\pm 1$, so as $g(X,U),g(Y,T),g(S,R)$. Following the same argument as the $C_2$ case, we have
\be
\nu_{3D}=g(\Gamma,Z)g(X,U)g(Y,T)g(S,R)
\ee
If the mirror plane is perpendicular to $\hat x+\hat y$ as in Fig.(\ref{bzc2})(d), $S_{\sigma T}$ consists of two lines $\{ \overline{SS''},\overline{RR''} \}$. The TRIM $\Gamma,S,Z,R$ are invariant under the mirror symmetry $\sigma$ but $X',Y,U',T$ are not. Following the same argument, $g(X',Y)$ and $g(U',T)$ are fixed to 1 because these two points interchange under $\sigma$, therefore they do not contribute to $\nu_{3D}$. In this case we have
\be
\nu_{3D}=g(\Gamma,S)g(Z,R)
\ee
In general, similar to the $C_2$ case above, we can define $L_{\sigma T}$ to be the set of straight paths inside $S_{\sigma T}$ such that each path $\gamma\in L_{\sigma T}$ connects two mirror-invariant TRIM, different paths in $L_{\sigma T}$ do not cross with each other, and each mirror-invariant TRIM is connected by one path in $L_{\sigma T}$. In Fig.(\ref{bzc2}) (c) and (d) $L_{\sigma T}$ reduces to $\{ \overline{\Gamma Z},\overline{XU},\overline{YT},\overline{SR} \}$ and $\{ \overline{\Gamma S},\overline{ZR} \}$ respectively. Note that in Fig.(\ref{bzc2})(d) we cannot take $\{ \overline{\Gamma Z},\overline{SR} \}$ because these lines are not inside $\sigma T$ invariant subspace. With this definition, the index is written as
\be
\nu_{3D}=\prod_{\gamma\in L_{\sigma T}} \textmd{Sign}[ g(\gamma)]
\label{gnumir}
\ee
Here the sign function takes account of the fact that $|g(\gamma)|$ can be smaller than 1 in a real calculation on discrete momentum points. With Eq.\eqref{gnuc2} and \eqref{gnumir}, we have developed a unified method to calculate FKM strong $Z_2$ topological index that requires examining only a 1D subspace of the BZ, for systems with two-fold rotation or mirror symmetry. If the system has multiple mirrors or $C_2$ axes, this method will work if we focus on any one of them.

It is also worthwhile to explore the case when the system has space inversion symmetry $I$ in addition to $T$. It turns out that our method is still applicable, although the $Z_2$ index can be determined directly by inversion eigenvalues. In this case Eq.\eqref{gsym} when applied to symmetry operator $IT$ shows that that $g(k_a,k_b)$ is real everywhere, since every momentum point is invariant under $IT$. This implies $g(K_i,K_j)$ will be quantized to $\pm 1$ for any pair of TRIM $K_i$ and $K_j$. Therefore our method is applicable again and we have $(-1)^{\nu_{3D}}=\prod_{\gamma} g(\gamma)$, where the product ranges over the four lines that connect the eight TRIM.

\section{Higher-order topological insulators with $C_4T$ symmetry}

\begin{figure}
\includegraphics[width=1.8 in]{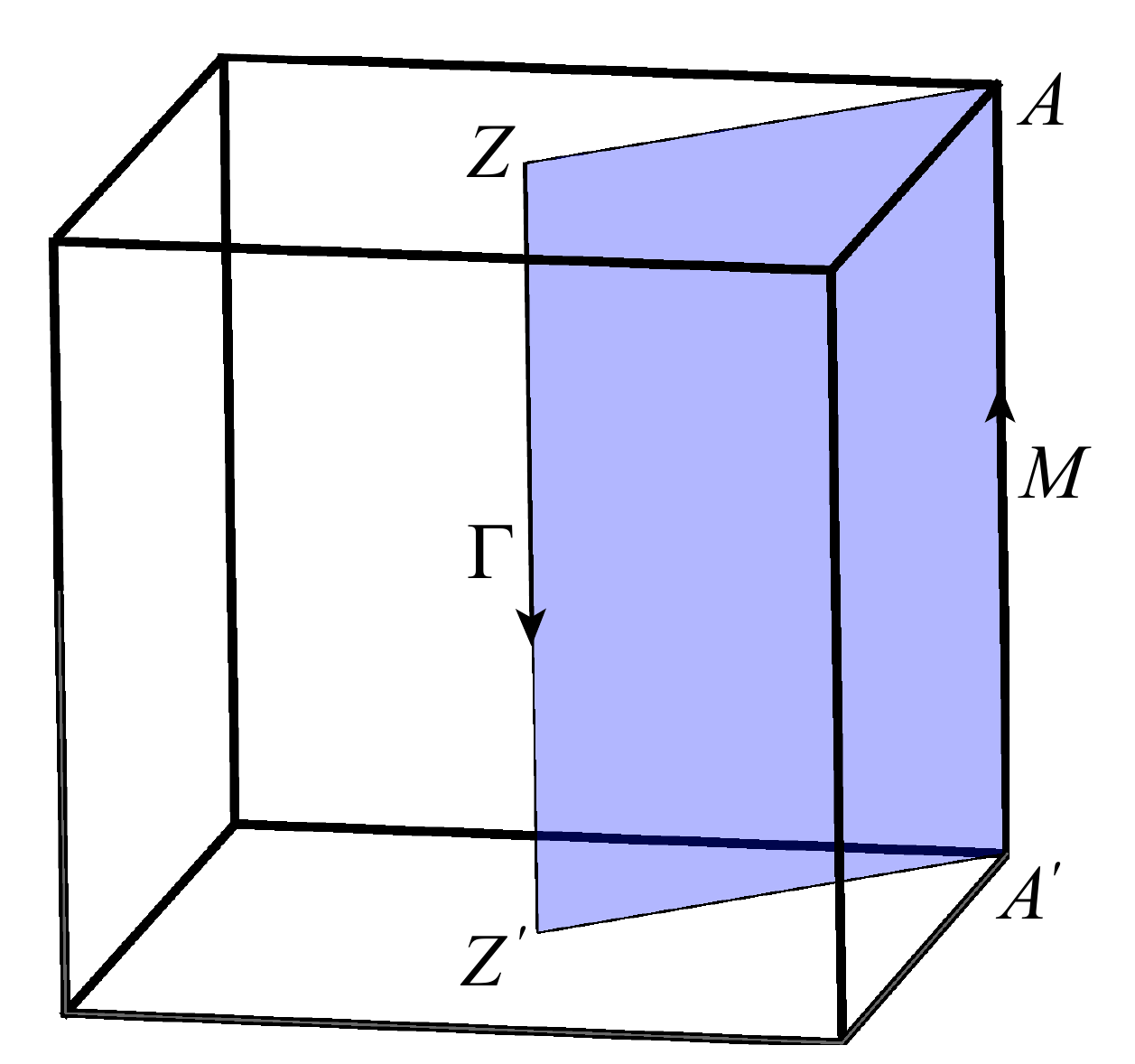}
\caption{ The plot of a 3D Brillouin zone for higher-order topological insulator protected by $C_4T$ symmetry. The $C_4$ axis is along $\hat z$ direction and $\tau$ is the colored area.  }
\label{bzHOTI}
\end{figure}

The Pfaffian formalism that describes the FKM $Z_2$ index for topological insulators has recently been generalized to higher-order topological insulators~\cite{pfHOTI}. We show that the line quantity $g(k_a,k_b)$ can be applied to calculate the topological index for higher-order topological insulators as well. Consider the chiral second-order topological insulator protected by $C_4T$ symmetry. The topological index is the magnetoelectric polarization $P_3$, which is quantized to $0$ or $\frac{1}{2}$ by $C_4T$ symmetry, satisfying a $Z_2$ classification. If $P_3=\frac{1}{2}$ the system is topological, with gapless states localized at the one-dimensional hinges, but the system remains gapped at the bulk and 2D side surfaces. Time-reversal symmetry in this system is broken, therefore the original FKM Pfaffian formalism which requires time-reversal symmetry is not applicable. However, it has been shown in Ref.\onlinecite{pfHOTI} that with the definition of a composite operator $\Theta=\frac{C_4T+C_4^{-1}T}{\sqrt{2}}$, the topological index $P_3$ can be represented by a Pfaffian formula in which $T$ is replaced by $\Theta$:
\bea
&&2P_3=\frac{1}{2\pi i} \oint_{\partial \tau}  dk \cdot \mathbf{\nabla} \log \operatorname{Pf}[M_{\Theta}(k)]\mod 2  \label{p3pf} \\
&&M_{\Theta,mn}(k)=\bra{u_m(k)} \Theta \ket{u_n(k)}
\eea
Here $\tau$ is the colored area in Fig.(\ref{bzHOTI}) and the $C_4$ axis is along $\hat z$ direction. Eq.\eqref{p3pf} has the same form as Eq.\eqref{pf2D} with $T$ replaced by $\Theta$. We can make the same replacement in the definition of $g(k_a,k_b)$ in Eq.\eqref{gwilsonl} to define:
\be
g_\Theta(k_a,k_b)=\frac{\operatorname{Pf}[M_\Theta(k_b)]}{\operatorname{Pf}[M_\Theta(k_a)]} \det \left[ \langle u_m(k_a)| \prod_{k_i\in \overline{k_a k_b}}^{k_a\leftarrow k_b} P_{k_i} | u_n(k_b)\rangle \right]
\ee
The line quantity $g_\Theta(k_a,k_b)$ defined in this way is gauge-invariant as well. Following the same procedure that lead to Eq.\eqref{gpfaffian}, we arrive at
\be
2P_3=\frac{1}{2\pi i} \int_{Z'}^Z   dk \cdot \mathbf{\nabla} \log \overline{g}_\Theta(k) \mod 2
\label{gp3}
\ee
Here $\overline{g}_\Theta(k)=g_\Theta(k,k+\pi \hat x+\pi\hat y)$ and the integral is along the straight path from $Z'$ to $Z$. Therefore the magnetoelectric polarization $P_3$ can be expressed in terms of an integral of the line quantity $g_\Theta(k_a,k_b)$. The advantage of this method is that it involves only gauge-independent objects, therefore a smooth gauge is not needed.

\section{Conclusion}

We show that in systems where topological indices cannot be reduced to high-symmetry-point symmetry eigenvalues, certain symmetries can still help simplify the calculation of the index. With the definition of the gauge-invariant line quantity $g(k_a,k_b)$, we present a unified way to calculate topological index by examining only 1D subspace of the Brillouin zone for systems with either $C_2$ or mirror symmetry in addition to time-reversal symmetry. Our method is applicable to a wide range of systems because among all the 32 point groups, 30 of them contain such a symmetry, except for $C_3$ the the trivial group $C_1$. This approach also finds its application in higher-order topological insulators.


%

\begin{widetext}

\appendix

\section{Properties of $g(\widetilde{k_a k_b})$}

Here we prove some equalities involving $g(\widetilde{k_a k_b})$ defined in Eq.(\ref{gwilson}). First we investigate the case when $k_a$ and $k_b$ are infinitely close to each other, denoted as $k_1$ and $k_2$. In this case we have $g(k_1,k_2)=\frac{\operatorname{Pf}[M(k_2)]}{\operatorname{Pf}[M(k_1)]}\det[ W(k_1,k_2) ]$ and $W(k_1,k_2)$ reduces to the overlap between the wave functions at $k_1$ and $k_2$: $W_{mn}(k_1,k_2)=\langle u_m(k_1)\ket{u_n(k_2)}$. $g(k_1,k_2)$ is invariant under gauge transformation $|u_m(k)\rangle\rightarrow U_{nm}(k)|u_n(k)\rangle$. Since under this transformation, $M_{mn}(k)=\langle u_m(k)| T|u_n(k) \rangle\rightarrow (U(k)^\dagger M(k) U(k)^*)_{mn}$, $W_{mn}(k_1,k_2)=\langle u_m(k_1)|u_n(k_2)\rangle \rightarrow (U(k_1)^\dagger W(k_1,k_2) U(k_2))_{mn}$, therefore
\bea
\operatorname{Pf}[M(k)]&\rightarrow& \operatorname{Pf}[M(k)] \det[U(k)]^*   \nonumber\\
\det[W(k_1,k_2)]&\rightarrow&\det[W(k_1,k_2)]\det[U(k_1)]^* \det[U(k_2)] \nonumber\\
g(k_1,k_2)&=&\frac{\operatorname{Pf}[M(k_2)]}{\operatorname{Pf}[M(k_1)]}\det[W(k_1,k_2)]\rightarrow g(k_1,k_2)
\eea
This completes the proof that $g(k_1,k_2)$ is gauge-invariant. For a general path $\widetilde{k_a k_b}$, divide the path by small segments $(k_i,k_{i+1})$ and by definition in Eq.\eqref{gwilson}, $g(\widetilde{k_a k_b})=\prod_i g(k_i,k_{i+1})$. For each small segment $g(k_i,k_{i+1})$ is gauge-invariant, therefore $g(\widetilde{k_a k_b})$ is gauge-invariant as well.

Next we prove Eq.(\ref{gsym}). Suppose the system has an anti-unitary symmetry $C$ such that it commutes with $T$ and it transforms momentum $k$ to $Ck$, define the sewing matrix of $C$ as
\be
R_{mn}(k)=\langle u_m(C k)|C| u_n(k)\rangle,
\ee
Insert identity $\textbf{1}= P_{occ}(k)+P_{unocc}(k)=\sum_{i\in occ} |u_i(k)\rangle\langle u_i(k)| +\sum_{i\in unocc} |u_i(k)\rangle\langle u_i(k)|$ to each $\textbf{1}$ in the identity below:
\be
\langle u_m(C k)|T |u_n(C k)\rangle=\langle u_m(C k)|C \textbf{1} T \textbf{1} C^{-1} |u_n(C k)\rangle,
\ee
where $P_{occ}$ and $P_{unocc}$ are projection to occupied and unoccupied bands respectively, $m$ and $n$ belong to occupied bands. Since $\langle u_m(C k)|C P_{unocc}(k)=0$, we can omit $P_{unocc}$ in the insertion and get
\be
\langle u_m(C k)|T |u_n(C k)\rangle=\sum_{i,j\in occ}\langle u_m(C k)|C |u_i(k)\rangle\langle u_i(k)| T |u_j(k)\rangle^* \langle u_j(k)| C^{-1} |u_n(C k)\rangle,
\label{insert}
\ee
where the conjugation is due to the fact that $T$ and $C$ are anti-unitary. Notice that $\langle u_j(k)| C^{-1} |u_n(C k)\rangle=\langle u_n(C k)| C |u_j(k)\rangle$, the above equation implies
\bea
M(C k)&=&R(k) M(k)^* R(k)^T \nonumber\\
\operatorname{Pf}[M(C k)]&=& \det[R(k)] \operatorname{Pf}[M(k)]^*
\eea
To prove Eq.(\ref{gsym}) we still need to compute $W(C k_1, C k_2)$. Using $|u_m(C k)\rangle = R^\dagger_{nm}(k)C|u_n(k)\rangle$ we get
\bea
W_{ij}(C k_1,C k_2)&=&R_{in}(k_1) \langle C u_n(k_1)|C u_m(k_2)\rangle R^\dagger_{mj}(k_2)\rangle  \nonumber\\
&=& R_{in}(k_1) \langle  u_n(k_1)| u_m(k_2)\rangle^* R^\dagger_{mj}(k_2)\rangle \nonumber\\
\det[W(C k_1,C k_2)]&=& \det[W(k_1,k_2)]^* \det[R(k_1)] \det[R(k_2)]^*
\label{wij}
\eea
Therefore when $C$ is anti-unitary we have:
\bea
g(C k_1,C k_2)=\frac{\operatorname{Pf}[M(C k_2)]}{\operatorname{Pf}[M(C k_1)]}\det[W(C k_1,C k_2)]=\frac{\operatorname{Pf}[M(k_2)]^*}{\operatorname{Pf}[M(k_1)]^*}\det[W(k_1,k_2)]^*=g(k_1,k_2)^*
\label{antic}
\eea

If $C$ is unitary instead, Eq.\eqref{insert} will be changed to
\be
\langle u_m(C k)|T |u_n(C k)\rangle=\sum_{i,j\in occ}\langle u_m(C k)|C |u_i(k)\rangle\langle u_i(k)| T |u_j(k)\rangle \langle u_j(k)| C^{-1} |u_n(C k)\rangle^*,
\ee
Using $\langle u_j(k)| C^{-1} |u_n(C k)\rangle^*=\langle u_n(C k)| C |u_j(k)\rangle$ for unitary symmetry $C$, we have
\bea
M(C k)&=&R(k) M(k) R(k)^T \nonumber\\
\operatorname{Pf}[M(C k)]&=& \det[R(k)] \operatorname{Pf}[M(k)]
\eea
For $W(C k_1,C k_2)$, Eq.\eqref{wij} will be
\bea
W_{ij}(C k_1,C k_2)&=&R_{in}(k_1) \langle C u_n(k_1)|C u_m(k_2)\rangle R^\dagger_{mj}(k_2)\rangle  \nonumber\\
&=& R_{in}(k_1) \langle  u_n(k_1)| u_m(k_2)\rangle R^\dagger_{mj}(k_2)\rangle \nonumber\\
\det[W(C k_1,C k_2)]&=& \det[W(k_1,k_2)] \det[R(k_1)] \det[R(k_2)]^*
\eea
Therefore if $C$ is unitary we have:
\bea
g(C k_1,C k_2)=\frac{\operatorname{Pf}[M(C k_2)]}{\operatorname{Pf}[M(C k_1)]}\det[W(C k_1,C k_2)]=\frac{\operatorname{Pf}[M(k_2)]}{\operatorname{Pf}[M(k_1)]}\det[W(k_1,k_2)]=g(k_1,k_2)
\label{unic}
\eea

A general path $\widetilde{k_a k_b}$ can be divided by small segments $(k_i,k_{i+1})$ so that $g(\widetilde{k_a k_b})=\prod_i g(k_i,k_{i+1})$. Eq.\eqref{gsym} is proved by applying Eq.\eqref{antic} or \eqref{unic} to each segment $g(k_i,k_{i+1})$.

\section{ Triviality of TRIM that are not invariant under $C_2$ or mirror symmetry }

In this section we give a more detailed proof of the assertion that the $Z_2$ index $\nu_{2D}$ in the time-reversal invariant plane passing through $X,Y,T,U$ in Fig.(\ref{bzc2})(b) and $X',Y,T,U'$ in Fig.(\ref{bzc2})(d) are trivial. These 2D planes are shown in Fig.(\ref{bzcut}), which are obtained from a cut in Fig.(\ref{bzc2})(b) and (d) respectively. This proof utilizes the interpretation of the line quantity $g(\widetilde{k_a k_b})$ as a measure of Pfaffian in the parallel transport gauge.

In Fig.(\ref{bzcut})(a) the system has a two-fold rotational symmetry $C_2$ perpendicular to the plane which is inherited from the 3D system. However, the $C_2$ rotation centers are located at the black dots that bisect two TRIM. This type of $C_2$ operator is different from the conventional two-fold rotation that can be realized by a 2D lattice in real space, since in that case the rotation center in the momentum space will always locate at some TRIM. If we choose the origin to be at $Y$ and denote the components of $k$ along $G_1$ and $G_2$ direction as $k_x$ and $k_y$ respectively, the $C_2$ operator at the midpoint of $\overline{XY}$ generates a transformation $(k_x,k_y)\rightarrow(G_1/2-k_x,-k_y)$, and time-reversal generates $(k_x,k_y)\rightarrow(-k_x,-k_y)$. Therefore the combined operation $C_2T$ gives $(k_x,k_y)\rightarrow(k_x-G_1/2,k_y)$. Define $\overline{g}(\vec k)=g(\vec k,\vec k+\vec G_2/2)$ for $\vec k\in\overline{XX_2}$, then from Eq.\eqref{gsym} the $C_2T$ symmetry requires
\bea
&&\overline{g}(\vec k)=\overline{g}(\vec k+\vec G_1/2)^* \nonumber\\
&&\textmd{Im} \log \overline{g}(\vec k)= -\textmd{Im} \log \overline{g}(\vec k+\vec G_1/2) ,\ \vec k\in\overline{XX_2}
\label{c2cancel}
\eea
Denote the colored region in Fig.(\ref{bzcut}) as $\tau$ and use the same derivation that lead to Eq.\eqref{gpfaffian}, the $Z_2$ index in this plane is
\bea
\nu_{2D}&=&\frac{1}{2\pi} \textmd{Im} \int_{X_2}^X   d\vec k \cdot \mathbf{\nabla} \log \overline{g}(\vec k) \nonumber\\
\overline{g}(\vec k)&=&g(\vec k,\vec k+\vec G_2/2)
\label{c2int}
\eea
From Eq.\eqref{c2cancel}, the integrand in Eq.\eqref{c2int} at $\vec k$ cancels that at $\vec k+\vec G_1/2$, which leads to $\nu_{2D}=0$. Therefore the plane has a trivial $Z_2$ index due to the $C_2$ symmetry.

In Fig.(\ref{bzcut})(b) the system has mirror planes inherited from the 3D system located at the thin vertical lines. The mirror plane to the right has the transformation $(k_x,k_y)\rightarrow(G_1/2-k_x,k_y)$. Define $\overline{g}(\vec k)=g(\vec k,\vec k+\vec G_2/2)$ for $\vec k\in\overline{X'X_3}$, from Eq.\eqref{gsym} the mirror symmetry requires
\be
\overline{g}(\vec k)=\overline{g}(\vec G_1/2-\vec k)
\label{mircancel}
\ee
The $Z_2$ index from Eq.\eqref{gpfaffian} is
\bea
\nu_{2D}&=&\frac{1}{2\pi} \textmd{Im} \int_{X'}^{X_3}   d\vec k \cdot \mathbf{\nabla} \log \overline{g}(\vec k)
\eea
Therefore Eq.\eqref{mircancel} requires the integrand at $\vec k$ to cancel that at $\vec G_1/2-\vec k$, leading to $\nu_{2D}=0$. Therefore the $Z_2$ index for this 2D plane is trivial due to the mirror symmetry.

\begin{figure}
\includegraphics[width=4.2 in]{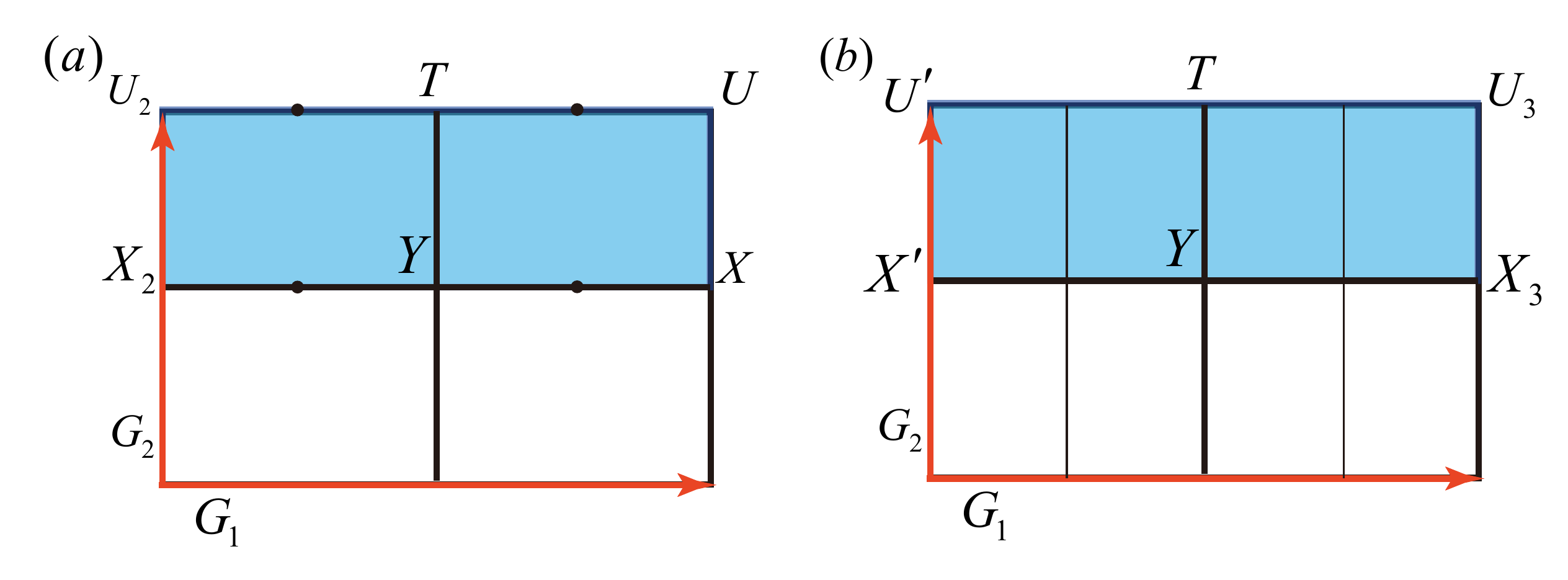}
\caption{ The plot of time-reversal invariant planes obtained from a cut in Fig.(\ref{bzc2})(b) and (d). We define the origin in these planes to be at $Y$. The black dots in (a) are two-fold rotation centers. The thin vertical lines in (b) are mirror planes. Note that none of these two-fold rotation centers or mirror planes pass through the origin at $Y$.  }
\label{bzcut}
\end{figure}


\end{widetext}

\end{document}